%% file: Lambda1520_PbPb_paper.tex
\documentclass[ALICE,manyauthors]{cernphprep}
\usepackage[comma,square,numbers,sort&compress]{natbib}
\usepackage{hyperref}
\usepackage{graphicx}
\usepackage{dcolumn}
\usepackage{bm}
\usepackage{amssymb}
\usepackage{amsfonts}
\usepackage{graphics}
\usepackage{grffile}
\usepackage{epsfig}
\usepackage{units}
\usepackage[usenames]{color}
\usepackage[normalem]{ulem}
\usepackage{color}
\usepackage[utf8]{inputenc}
\usepackage[T1]{fontenc}
\usepackage{subfigure}
\usepackage{lineno}
\usepackage{tabularx,booktabs}

\input{commands.tex}

\begin{document}

\begin{titlepage}

\PHyear{2018}
\PHnumber{116}      % required, will be obtained from PH
\PHdate{9 May}  % required, will be obtained from PH

  \title{Suppression of \Lambdastar resonance production\\in central \PbPb collisions at \sqrtsNN~=~2.76 TeV}
  \ShortTitle{Suppression of \Lambdastar production in central \PbPb collisions}
  
  \Collaboration{ALICE Collaboration
    \thanks{See Appendix~\ref{app:collab} for the list of collaboration members}}
  \ShortAuthor{ALICE Collaboration}

  \begin{abstract}
    \input{abstract}
  \end{abstract}
  
\end{titlepage}

\input{paper}

\newenvironment{acknowledgement}{\relax}{\relax}
\begin{acknowledgement}
  \section*{Acknowledgements}
  \input{fa_2018-05-08.tex}
\end{acknowledgement}

\bibliographystyle{utphys}
\bibliography{biblio}{}

\newpage

\appendix
\section{The ALICE Collaboration}
\label{app:collab}
\input{Alice_Authorlist_2018-Apr-24.tex}

\end{document}

%% file: commands.tex
\usepackage{xspace}
\usepackage{color}
\usepackage{xcolor}
\usepackage{rotating}
\definecolor{light-gray}{gray}{0.8}

\newcommand{\cmnt}[1]{}

\newcommand{\pt}{\ensuremath{p_{\rm T}}\xspace}
\newcommand{\mt}{\ensuremath{m_{\rm T}}\xspace}

\newcommand{\sqrtsNN}{\ensuremath{\sqrt{s_{\rm NN}}}\xspace}

\newcommand{\dndy}{\ensuremath{{\rm d}N/{\rm d}y}\xspace}

\newcommand{\dnchdetacubic}{\ensuremath{\average{{\rm d}N_{\rm ch}/{\rm d}\eta}^{1/3}}\xspace}

\newcommand{\etaless}[1]{\ensuremath{\left|\eta\right| < #1}\xspace}
\newcommand{\yless}[1]{\ensuremath{\left|y\right| < #1}\xspace}

\newcommand{\MeVc}{MeV/$c$\xspace}
\newcommand{\GeVc}{GeV/$c$\xspace}

\newcommand{\fmc}{fm/$c$\xspace}
\newcommand{\pp}{\ensuremath{\rm pp}\xspace}

\newcommand{\PbPb}{Pb--Pb\xspace}
\newcommand{\dAu}{d--Au\xspace}
\newcommand{\AuAu}{Au--Au\xspace}

\newcommand{\average}[1]{\ensuremath{\langle #1 \rangle}\xspace}
\newcommand{\dedx}{\ensuremath{{\rm d}E/{\rm d}x}\xspace}
\newcommand{\muB}{\ensuremath{\mu_{\rm B}}\xspace}
\newcommand{\Tch}{\ensuremath{T_{\rm ch}}\xspace}

\newcommand{\pKminus}{\ensuremath{{\rm K}^{-}}\xspace}

\newcommand{\pProtonNS}{\ensuremath{p}}

\newcommand{\pLambda}{\ensuremath{\Lambda}\xspace}

\newcommand{\apLambda}{\ensuremath{\overline{\Lambda}}\xspace}

\newcommand{\Lambdastar}{\ensuremath{\Lambda(1520)}\xspace}

\newcommand{\Kstar}{\ensuremath{{\rm K}^{\ast}(892)^{0}}\xspace}

\newcommand{\Lambdastardecay}{\ensuremath{\Lambda(1520) \rightarrow \pProtonNS \pKminus}\xspace}

%% file: abstract.tex
The production yield of the \Lambdastar baryon resonance is measured at mid-rapidity in \PbPb collisions at \sqrtsNN~=~2.76 TeV with the ALICE detector at the LHC.
The measurement is performed in the \Lambdastardecay (and charge conjugate) hadronic decay channel as a function of the transverse momentum (\pt) and collision centrality.
The ratio of the \pt-integrated production of \Lambdastar baryons relative to \pLambda baryons in central collisions is suppressed by about a factor of 2 with respect to peripheral collisions.
This is the first observation of the suppression of a baryonic resonance at the LHC and the first $3\sigma$ evidence of \Lambdastar suppression within a single collision system.
The measured \Lambdastar/\pLambda ratio in central collisions is smaller than the value predicted by the statistical hadronisation model calculations.
The shape of the measured \pt distribution and the centrality dependence of the suppression are reproduced by the EPOS3 Monte Carlo event generator.
The measurement adds further support to the formation of a dense hadronic phase in the final stages of the evolution of the fireball created in heavy-ion collisions, lasting long enough to cause a significant reduction in the observable yield of short-lived resonances.

%% file: paper.tex
\input{introduction}

\input{detector}

\input{analysis}
\input{systematics}
\input{results}
\input{conclusions}

%% file: introduction.tex
\section{Introduction}
High-energy heavy-ion collisions provide an excellent means to study the properties of nuclear matter under extreme conditions and the phase transition to a deconfined state of quarks and gluons (Quark-Gluon Plasma, QGP~\cite{Cabibbo:1975ig}) predicted by lattice QCD calculations~\cite{Laermann:2003cv}.
The bulk properties of the matter created in high-energy nuclear reactions have been widely studied at the Relativistic Heavy Ion Collider (RHIC) and at the Large Hadron Collider (LHC), and are well described by hydrodynamic and statistical models.
The initial hot and dense partonic matter rapidly expands and cools, eventually undergoing a transition from the QGP to a hadron gas phase~\cite{Muller:2006ee}.
The relative abundances of stable particles are consistent with chemical equilibrium and are successfully described by Statistical Hadronisation Models (SHMs)~\cite{Andronic:2005yp,Becattini:2009fv,Cleymans:1998fq}.
They are determined by the ``chemical freeze-out'' temperature \Tch and the baryochemical potential \muB~\cite{Andronic:2005yp,Andronic:2011yq}, reflecting the thermodynamic characteristics of the so-called ``chemical freeze-out''.
In the final stage of the collision a dense hadron gas is expected to form and to expand until the system eventually decouples when elastic interactions cease.
Hadronic resonances with lifetimes shorter than or comparable to the timescale of the fireball evolution (a few \fmc) are sensitive probes of the dynamics and properties of the medium formed after hadronisation~\cite{Markert:2008jc}.
Within the SHM, they are expected to be produced with abundances consistent with the chemical equilibrium parameters \Tch and \muB, but the measured yields might be modified after the chemical freeze-out by the hadronic phase.
Due to their short lifetimes, resonances can decay within the hadronic medium, which can alter or destroy the correlation among the decay daughters via interactions (re-scattering) with the surrounding hadrons, hence reducing the observed yield. 
Alternatively, an increase (re-generation) might also be possible due to resonance formation in the hadronic phase~\cite{Bleicher:2002dm}. 
The observed yield of hadronic resonances depends on the resonance lifetime, the duration of the hadronic phase and the relative scattering cross-section of the decay daughters within the hadronic medium.

Recent results from the ALICE Collaboration in \PbPb collisions at \sqrtsNN = 2.76 TeV~\cite{Abelev:2014uua} show that the production yields of the \Kstar resonance are suppressed in central collisions with respect to peripheral collisions and are overestimated by SHM predictions. 
This phenomenon might be due to the short \Kstar lifetime ($\tau~\sim$~4~\fmc) and suggests the dominance of destructive re-scattering over re-generation processes in the hadronic phase.
No suppression is observed for the longer-lived $\phi$(1020) meson ($\tau~\sim$~46~\fmc), indicating that it decays mostly outside the fireball.
Both observations are in agreement with the calculations of EPOS3, a model that includes a microscopic description of the hadronic phase~\cite{Knospe:2015nva}.
Within this model, the lifetime of the hadronic phase formed in central \PbPb collisions at \sqrtsNN = 2.76 TeV is predicted to be $\sim$~10~\fmc.
The measurement of the production of the \Lambdastar baryonic resonance, owing to its characteristic lifetime ($\tau~\sim$~12.6~\fmc), serves as an excellent probe to further constrain the formation, the evolution and the characteristics of the hadronic phase. 

In this article, we present the first measurement of \Lambdastar production in \PbPb collisions at \sqrtsNN = 2.76 TeV.
The measurement is performed at midrapidity, \yless{0.5}, with the ALICE detector~\cite{Aamodt:2008zz} at the LHC and is based on an analysis of about 15 million minimum-bias \PbPb collisions recorded in 2010.
Previous results on \Lambdastar production in high-energy hadronic collisions have been reported by STAR in \pp, \dAu and \AuAu collisions at \sqrtsNN = 200 GeV at RHIC~\cite{Adams:2006yu,Abelev:2008yz}.

%% file: detector.tex
\section{Experimental setup and data analysis}
A detailed description of the ALICE experimental apparatus and its performance can be found in~\cite{Abelev:2014ffa}.
The relevant features of the main detectors utilised in this analysis are outlined here.
The V0 detector is composed of two scintillator hodoscopes, placed on either side of the interaction point and covers the pseudorapidity regions $2.8 < \eta < 5.1$ and $-3.7 < \eta < -1.7$, respectively.
It is employed for triggering, background suppression and collision-centrality determination.
The Inner Tracking System (ITS) and the Time-Projection Chamber (TPC) provide vertex reconstruction and charged-particle tracking in the central barrel, within a solenoidal magnetic field of 0.5 T.
The ITS is a high-resolution tracker made of six cylindrical layers of silicon detectors.
The TPC is a large cylindrical drift detector of radial and longitudinal dimensions of about $85 < r < 247$ cm and $-250 < z < 250$ cm, respectively.
Charged-hadron identification is performed by the TPC via specific ionisation energy-loss (\dedx) and by the Time-Of-Flight (TOF) detector.
The TOF is located at a radius of 370-399 cm and measures the particle time-of-flight with a resolution of about 80 ps, allowing hadron identification at higher momenta.
A minimum-bias trigger was configured to select hadronic events with high efficiency, requiring a combination of hits in the two innermost layers of the ITS and in the V0 detector.
The contamination from beam-induced background is removed offline, as discussed in detail in~\cite{Aamodt:2010pb,Aamodt:2010cz}.
The collision centrality is determined based on the signal amplitude of the V0 detector, whose response is proportional to the event multiplicity.
A complete description of the event selection and centrality determination can be found in~\cite{Abelev:2013qoq}.

%% file: analysis.tex
\begin{figure}
\begin{center}
\includegraphics[width=\linewidth]{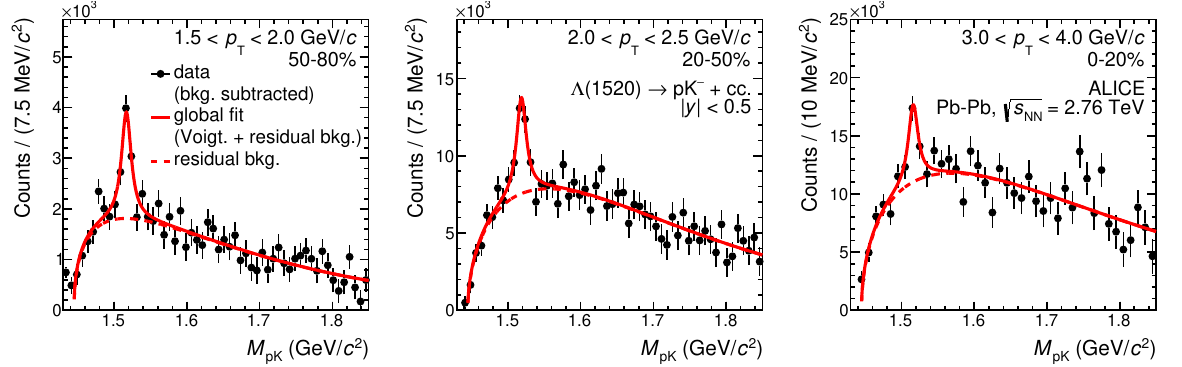}
\caption{Example invariant-mass distributions of the \Lambdastardecay (and charged conjugate) reconstruction after subtraction of the mixed-event background.
  The solid line represents the global fit (signal + residual background) to the data while the dashed line indicates the estimated residual background.
  The error bars indicate the statistical uncertainties of the data.}
\label{fig:invmass}
\end{center}
\end{figure}

The \Lambdastar resonance is reconstructed via invariant-mass analysis of its decay daughters in the hadronic decay channel \Lambdastardecay (and charge conjugate, cc.), with a branching ratio of 22.5 $\pm$ 0.5\% \cite{Olive:2016xmw}.
Particle and antiparticle states are combined to enhance the statistical significance of the reconstructed signal.
\Lambdastar refers to their sum in the following, unless otherwise specified.
Candidate daughters are selected from tracks reconstructed by the ITS and TPC, are required to have \pt $>$ 150 \MeVc and are restricted to the pseudorapidity range \etaless{0.8} for uniform acceptance and efficiency performance.
Furthermore, a cut on the impact parameter to the primary vertex is applied to reduce contamination from secondary tracks emanating from weak decays or from interactions with the detector material.
Details on the track selection can be found in~\cite{Abelev:2014uua}.
Kaons and protons are identified from the combined information of the track \dedx in the TPC and the time-of-flight measured by the TOF.
The invariant-mass distribution of unlike-sign pairs of selected kaon and proton tracks is constructed for each centrality class and \pt interval.
The rapidity of the candidate \Lambdastar is required to be within \yless{0.5}.
A large source of background from random combinations of uncorrelated hadrons affects the invariant-mass spectrum.
A mixed-event technique~\cite{LHote:1992jct} is employed to estimate the combinatorial background, using unlike-sign proton and kaon tracks taken from different events with similar characteristics.
The background is normalised and corrected for event-mixing distortions using a fit to the mixed/same-event ratio of like-sign pairs.
Figure~\ref{fig:invmass} shows examples of the reconstructed invariant-mass distribution after background subtraction.
The \Lambdastar raw yield is then extracted by means of a global fit, where a Voigtian function (the convolution of the non-relativistic Breit-Wigner with the Gaussian detector resolution) is used to describe the signal. The shape of the residual background resembles that of a Maxwell-Boltzmann distribution, therefore the residual background is fitted with a similar functional form
\begin{equation}
f_{\rm background}(m_{p\rm{K}}) = B \, \sqrt{(m_{p\rm{K}} - m_{\rm cutoff})^{n}} \, C^{3/2} \, \exp\left[-C \, (m_{p\rm{K}} - m_{\rm cutoff})^{n}\right], \nonumber
\end{equation}
where $B$ (normalization), $m_{\rm cutoff}$ (low-mass cut-off), $C$, and $n$ are free parameters.

The raw yields are corrected for the decay branching fraction and for detector acceptance, reconstruction, track-selection and particle-identification efficiency, evaluated through a detailed Monte Carlo simulation of the ALICE detector.
Simulation events are produced using the HIJING Monte Carlo event generator~\cite{Wang:1991hta} with the addition of \Lambdastar signals (particle and antiparticle states).
Particle transport is performed by GEANT3~\cite{Brun:1994aa}.

%% file: systematics.tex
\input{tables/systematics}

The main sources of systematic uncertainty on the corrected yields are summarised in Table~\ref{tab:systematics}.
They include the signal extraction procedures as well as the contributions related to the efficiency corrections (true \pt distribution of \Lambdastar, track selection, particle identification, material budget and hadronic cross-section) and event normalisation.
A significant fraction of this uncertainty, estimated to be about 12\%, is common to all centrality classes.

%% file: tables/systematics.tex
\begin{table}[b]
  \centering
  \begin{tabular*}{\textwidth}{@{\extracolsep{\fill}}lccc}
    \toprule
    & 0--20\% & 20--50\% & 50-80\% \\
    \midrule
    Signal extraction & 12 & 11 & 9 \\
    Global tracking efficiency & 10 - 10.5 & 10 - 10.5 & 10 - 10.5 \\
    TOF matching efficiency & 1 - 6.5 & 1 - 6.5 & 0 - 6.5 \\
    Particle identification & 3 & 3 & 3  \\
    Material and interactions & 3.5 - 2.5 & 3.5 - 2.5 & 4.5 - 2.5 \\
    \Lambdastar true \pt distribution & 3.5 - 1 & 4.5 - 1 & 2.5 - 1 \\
    Normalisation & 0.5 & 1.5 & 4.5 \\
    \midrule
    Total & 17 - 17.5 & 16.5 - 17 & 15.5 - 16.5\\
    \midrule
    Uncorrelated & 12 - 11.5 & 11.5 - 10.5 & 9.5 - 9.5 \\
    \bottomrule
  \end{tabular*}
  \caption{Main contributions to the systematic uncertainty of the \Lambdastar \pt-differential yield in 0--20\%, 20--50\%, and 50--80\% centrality classes.  The values are relative uncertainties (standard deviations expressed in \%). When appropriate, they are reported for the lowest and highest measured \pt bin. The total systematic uncertainty and the contribution uncorrelated across centralities (after removing the common uncertainties) are also reported.}
  \label{tab:systematics}
\end{table}

%% file: results.tex
\section{Results and discussion}
\begin{figure}[t]
\begin{center}
\includegraphics[width=\linewidth]{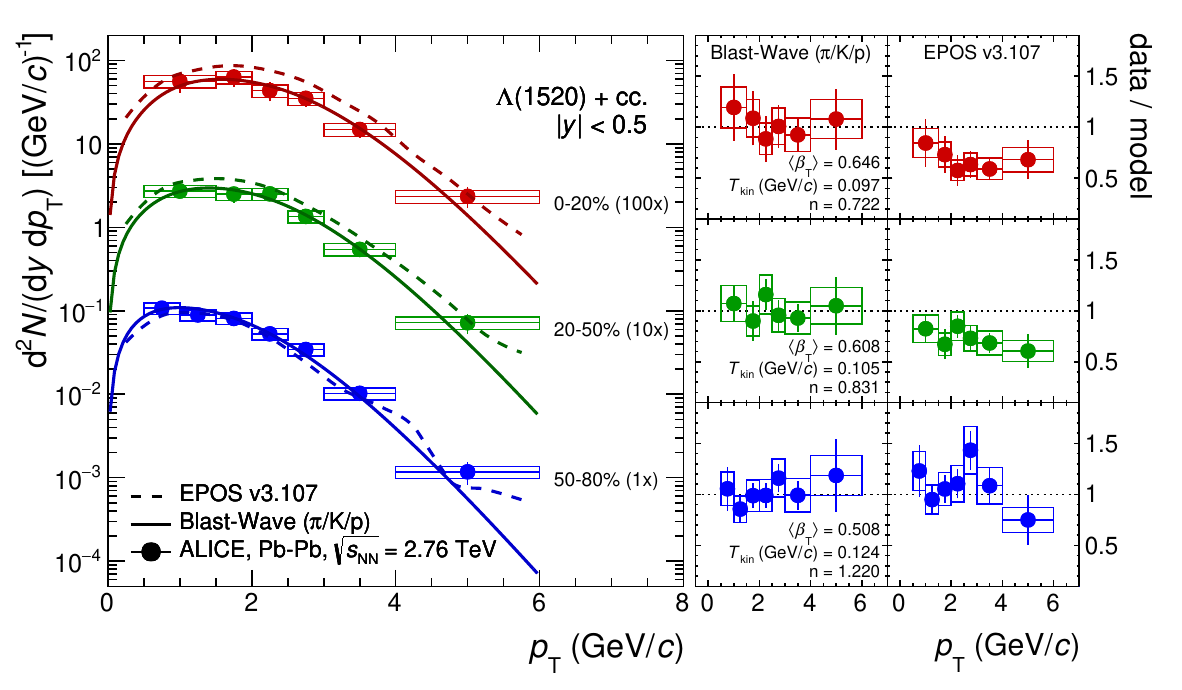}
\caption{(left) \pt-differential yields of \Lambdastar at midrapidity, \yless{0.5}, in the centrality classes 0--20\%, 20--50\%, and 50--80\%.
  The solid and dashed curves represent predictions from the Blast-Wave model (normalisation fitted to the data) and EPOS3, respectively. 
  The horizontal error bars represents the width of the measured \pt interval. 
  (right) Ratio of the data to the Blast-Wave and EPOS3 predictions.}
\label{fig:spectra}
\end{center}
\end{figure}

\input{tables/results}

\begin{figure}[t]
\begin{minipage}[t]{.4885\textwidth}
\centering
\includegraphics[width=1.0\textwidth]{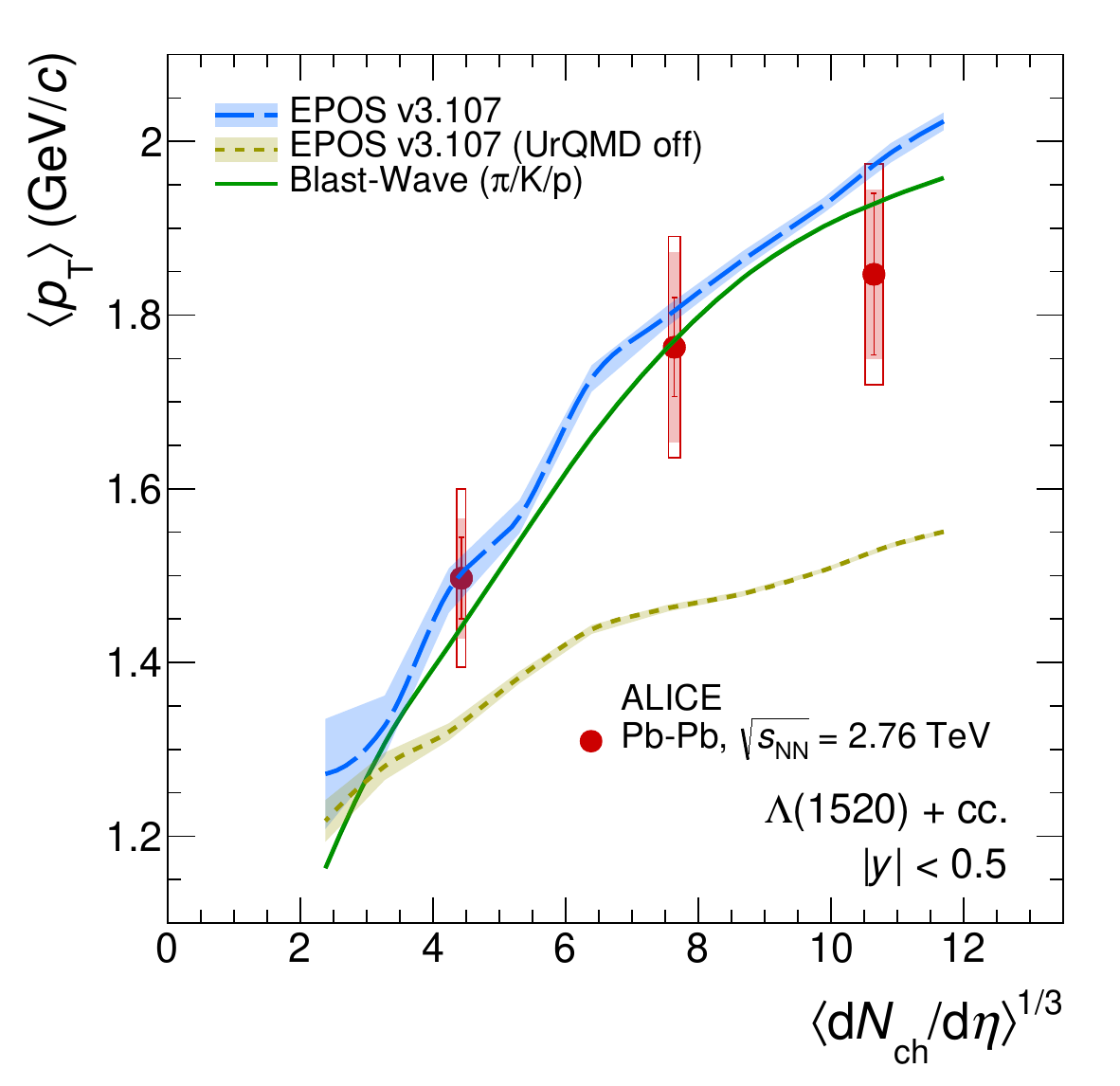}
\caption{\average{\pt} of \Lambdastar as a function of \dnchdetacubic.
  Statistical and systematic uncertainties are shown as bars and boxes, respectively.
  The solid line shows the Blast-Wave predictions.
  The dashed lines show the predictions from EPOS3 with and without the hadronic phase (UrQMD off).}
\label{fig:meanpt}
\end{minipage}
\hfill
\begin{minipage}[t]{.4885\textwidth}
\centering
\includegraphics[width=1.0\textwidth]{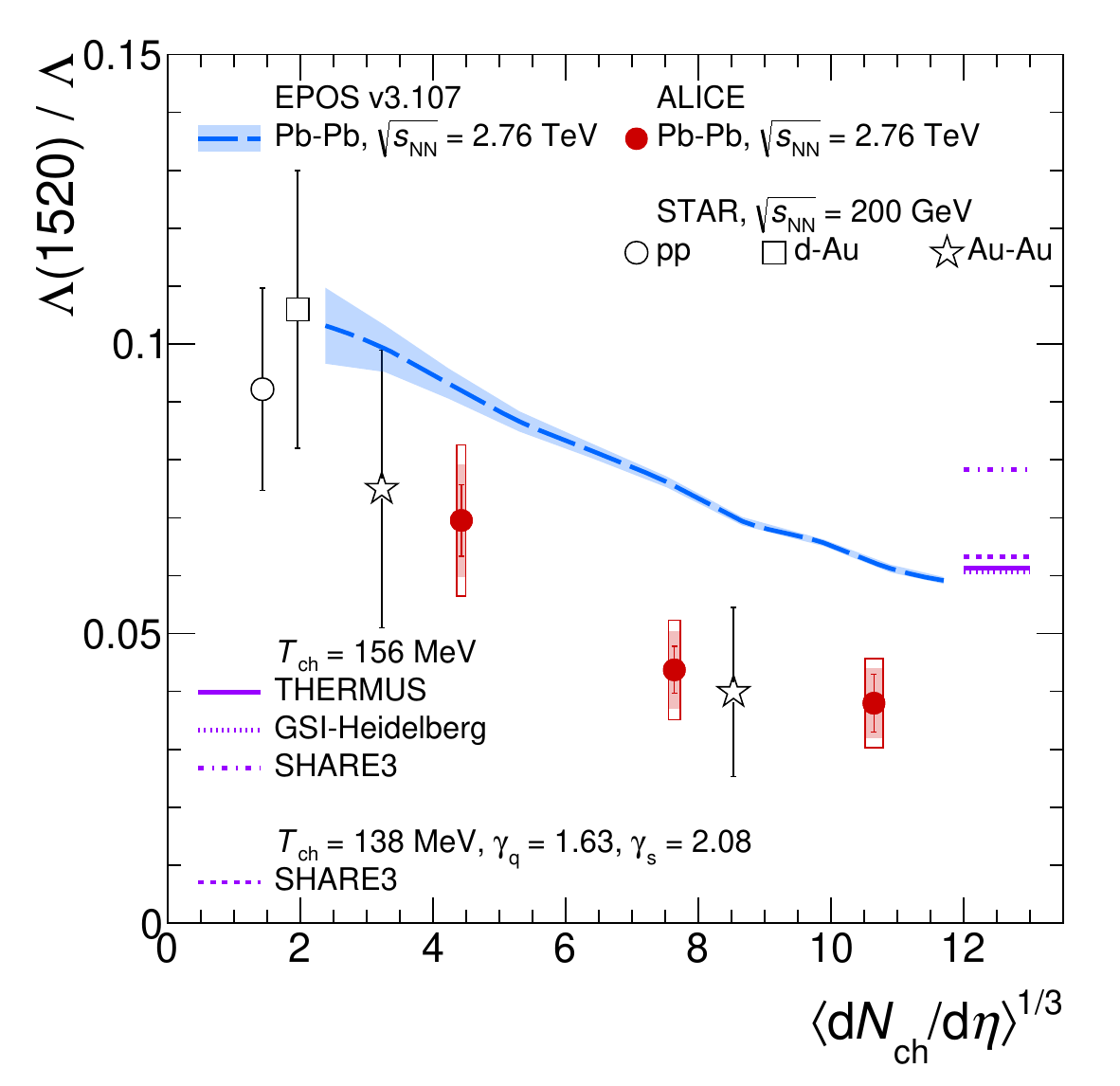}
\caption{\pt-integrated ratio of \Lambdastar/$\Lambda$ production as a function of \dnchdetacubic.
  Predictions from several SHMs and from EPOS3 are also shown.}
\label{fig:ratio}
\end{minipage}
\end{figure}

The fully-corrected \pt-differential yields of \Lambdastar measured in \yless{0.5} are shown in Fig.~\ref{fig:spectra} in the centrality classes 0--20\%, 20--50\%, and 50--80\%. 
The spectral shapes are compared with predictions from the Blast-Wave model~\cite{Schnedermann:1993ws}, which assumes particle production from thermal sources expanding with a common transverse velocity.
The parameters of the model are the ones obtained from published results on pion, kaon and proton production in \PbPb collisions~\cite{Abelev:2013vea}.
The good agreement of the Blast-Wave predictions with the data is consistent with the scenario where \Lambdastar undergoes a similar hydrodynamic evolution as pions, kaons and protons with a common transverse expansion velocity that increases with centrality.
The \pt distributions are also compared to predictions of the EPOS3 model~\cite{Knospe:2015nva}, a Monte Carlo generator founded on parton-based Gribov-Regge theory, which describes the full evolution of a heavy-ion collision.
The model employs viscous hydrodynamic calculations for the description of the expansion of the bulk partonic matter.
EPOS3 incorporates the Ultrarelativistic Quantum Molecular Dynamics (UrQMD)~\cite{Bass:1998ca,Bleicher:1999xi} transport model to describe the interactions among particles in the hadronic phase in a microscopic approach.
The results from the model are in rather good agreement with the measured \Lambdastar spectral shapes in all centrality classes, but the model overestimates the yields in central (0--20\%) and semi-central (20--50\%) collisions.

The \pt-integrated yield, \dndy, and the average transverse momentum, \average{\pt}, are computed by integrating the data and using extrapolations to estimate the yields in the unmeasured regions.
The extrapolations are obtained using the best fit of the Blast-Wave function to the \pt distributions.
Several other fit functions (Maxwell-Boltzmann, Fermi-Dirac, \mt-exponential, \pt-exponential) are employed to estimate the systematic uncertainties.
The fraction of the yields in the extrapolated regions are 6.2\%, 6.2\%, and 10.4\% for 0--20\%, 20--50\%, and 50--80\% centrality events, respectively.
The total \dndy systematic uncertainties are 17.2\%, 16.5\%, and 15.6\%, with a significant contribution common to all centrality classes of about 12\%.
The total systematic uncertainties on \average{\pt} are 6.9\%, 7.2\%, and 6.9\%. The values of \dndy and \average{\pt} for \Lambdastar are reported in Table~\ref{tab:results}.

The ratio of the \pt-integrated yield of \Lambdastar to that of its stable counterpart, \pLambda, highlights the characteristics of resonance production directly related to the particle lifetime, as possible effects due to valence-quark composition cancel.
The yields of \pLambda have been previously measured by ALICE in \PbPb collisions at \sqrtsNN~=~2.76~TeV~\cite{Abelev:2013xaa}. Since the centrality classes for \Lambdastar are different from those of the measured \pLambda yields, the latter have been interpolated from the measured values fitting their dependence on the number of participating nucleons in the collisions ($N_{\rm part}$) with the empirical parametrisation $a + b \langle N_{\rm part}\rangle^{c}$, where $a$, $b$, and $c$ are free parameters.
Moreover, the \apLambda yield, which was not published in~\cite{Abelev:2013xaa}, has been assumed equal to that of \pLambda, as expected at LHC energies~\cite{Abelev:2013vea}.
In the following, \pLambda refers to the sum of particle and antiparticle states, except for ALICE where it is defined as 2\pLambda.
The yield ratios \Lambdastar/\pLambda are reported in Table~\ref{tab:results}.

Figure~\ref{fig:meanpt} and~\ref{fig:ratio}, respectively, present the \average{\pt} of \Lambdastar and the \Lambdastar/\pLambda yield ratios as a function of the cubic root of the charged-particle multiplicity density at mid-rapidity~\cite{Aamodt:2010cz}, \dnchdetacubic.
The latter is used as a proxy for the system radius to emphasise the system-size dependence of \Lambdastar production, as suggested by femtoscopic studies using Bose-Einstein correlations~\cite{Aamodt:2011mr}.
The \average{\pt} increases significantly with increasing charged-particle multiplicity, hence with increasing collision centrality (Fig.~\ref{fig:meanpt}).
The value in central (0--20\%) collisions is 23\% higher than the one in peripheral (50--80\%) collisions.  
The measured \average{\pt} is compared to the prediction from the Blast-Wave model discussed previously~\cite{Schnedermann:1993ws,Abelev:2013vea}.
The good agreement confirms the consistency of the \Lambdastar data with a common hydrodynamic evolution picture.
The \average{\pt} results also highlight the good agreement with the prediction from the EPOS3~\cite{Knospe:2015nva} model.
It is noteworthy that EPOS3 fails to describe the data when the UrQMD transport stage is disabled, underlining the importance of the latter in the description of the evolution of heavy-ion collisions.

A gradual decrease of the \Lambdastar/\pLambda yield ratio with increasing charged-particle multiplicity is observed from peripheral to central \PbPb collisions (Fig.~\ref{fig:ratio}).
The \Lambdastar suppression in central \PbPb events with respect to peripheral events is measured as the double ratio
\begin{equation}
R^{\pLambda^{\ast}/\pLambda}_{\rm cp} = \frac{\left[\Lambdastar/\Lambda\right]_{\rm \, 0-20\% \,\,\,\,\, }}{\left[\Lambdastar/\Lambda\right]_{\rm \, 50-80\%}} = 0.54 \pm 0.08~({\rm stat}) \pm 0.12~({\rm syst}), \nonumber
\end{equation}
where common uncertainties cancel.
\Lambdastar/\pLambda in central collisions is about 45\% lower than in peripheral collisions.
The result provides the first evidence for \Lambdastar suppression in heavy-ion collisions, with a 3.1$\sigma$ confidence level.
The ratio is compared to grand-canonical equilibrium predictions from the GSI-Heidelberg~\cite{Andronic:2005yp}, THERMUS~\cite{Wheaton:2004vg} and SHARE3~\cite{Petran:2013dva} models, whose parameters have been determined from fits to stable particles~\cite{Floris:2014pta}.
The ratio is also compared to the non-equilibrium configuration implemented in SHARE3, where the under(over)-saturation parameters $\gamma_{s}$ (strange) and $\gamma_{q}$ (light quarks) are free~\cite{Petran:2013lja}.
All models describe the yield of stable hadrons well.
\Lambdastar/\pLambda in central collisions is lower than SHM predictions by values ranging from 37\% to 52\%, depending on the reference model.
Figure~\ref{fig:ratio} also shows the data from the STAR Collaboration at RHIC in \AuAu, \dAu, and pp collisions at \sqrtsNN~=~200 GeV~\cite{Adams:2006yu,Abelev:2008yz}.
The trend of the suppression is similar to the one seen from STAR data in \AuAu collisions at \sqrtsNN~=~200 GeV.
The current measurement of \Lambdastar suppression has a higher precision at 3.1$\sigma$ confidence level, as compared to the 1.8$\sigma$ confidence level of STAR data in \AuAu collisions.
Finally, the multiplicity-dependence of the \Lambdastar/\pLambda ratio is compared with the prediction from EPOS3~\cite{Knospe:2015nva} (Fig.~\ref{fig:ratio}).
It is important to note that the model, although it systematically overestimates the data, describes the trend of the suppression well.
The double ratio $R^{\pLambda^{\ast}/\pLambda}_{\rm cp}$ is in agreement with the data within the uncertainties, although the model is about 25\% higher.
This might hint to a longer lifetime of the hadronic phase than the value obtained from EPOS3 calculations ($\tau_{\rm hadr} \sim$ 8.5 \fmc) or to an imprecise description of the relevant hadronic cross-sections in the transport phase.
These observations highlight the relevance of the hadronic phase in the study of heavy-ion collisions and the importance of a microscopic description of the late hadronic interactions.

%% file: tables/results.tex
\begin{table}[b]
  \centering
  \begin{tabular*}{\textwidth}{@{\extracolsep{\fill}}rccc}
    \toprule
     & \dndy &  \Lambdastar/\pLambda & \average{\pt} (\GeVc) \\
    \midrule
     0--20\% & 1.56 $\pm$ 0.20 $\pm$ 0.27 (0.19) & 0.038 $\pm$ 0.005 $\pm$ 0.008 (0.006) & 1.85 $\pm$ 0.09 $\pm$ 0.13 (0.10) \\
    20--50\% & 0.70 $\pm$ 0.06 $\pm$ 0.12 (0.08) & 0.044 $\pm$ 0.004 $\pm$ 0.009 (0.007) & 1.76 $\pm$ 0.06 $\pm$ 0.13 (0.11) \\
    50--80\% & 0.22 $\pm$ 0.02 $\pm$ 0.03 (0.02) & 0.069 $\pm$ 0.006 $\pm$ 0.013 (0.010) & 1.50 $\pm$ 0.05 $\pm$ 0.10 (0.07) \\
\bottomrule
  \end{tabular*}
  \caption{\Lambdastar integrated yields, \pt-integrated ratio of \Lambdastar/$\Lambda$, \average{\pt} of \Lambdastar production and corresponding uncertainties in 0--20\%, 20--50\%, and 50--80\% centrality classes. The first and second uncertainties indicate the statistical and total systematic error, respectively. The values in parenthesis show the systematic uncertainty excluding the contributions common to all centrality classes.} 
  \label{tab:results}
\end{table}

%% file: conclusions.tex
\section{Conclusion}
In conclusion, the first measurement of \Lambdastar production in \PbPb collisions at \sqrtsNN = 2.76 TeV at the LHC has been presented.
The spectral shapes and \average{\pt} are consistent with the hydrodynamic evolution picture that describes pions, kaons and protons, indicating that the \Lambdastar experiences the same collective radial expansion, with a common transverse velocity which increases with collision centrality.
The comparison of the \average{\pt} results to EPOS3 predictions highlights the relevance of the hadronic phase in the study of heavy-ion collisions and the importance of a microscopic description of the late hadronic interactions.
The \pt-integrated ratio \Lambdastar/\pLambda is suppressed in central \PbPb collisions with respect to peripheral \PbPb collisions (first such evidence in heavy-ion collisions) and is lower than the value predicted by statistical hadronisation models.
The measurement adds further support to the formation of a dense hadronic phase in the latest stages of the evolution of the fireball created in high-energy heavy-ion collisions, lasting long enough to cause a significant reduction in the observable yield of short-lived resonances.

%% file: fa_2018-05-08.tex
% Version: 2018-05-08

The ALICE Collaboration would like to thank all its engineers and technicians for their invaluable contributions to the construction of the experiment and the CERN accelerator teams for the outstanding performance of the LHC complex.
The ALICE Collaboration gratefully acknowledges the resources and support provided by all Grid centres and the Worldwide LHC Computing Grid (WLCG) collaboration.
The ALICE Collaboration acknowledges the following funding agencies for their support in building and running the ALICE detector:
A. I. Alikhanyan National Science Laboratory (Yerevan Physics Institute) Foundation (ANSL), State Committee of Science and World Federation of Scientists (WFS), Armenia;
Austrian Academy of Sciences and Nationalstiftung f\"{u}r Forschung, Technologie und Entwicklung, Austria;
Ministry of Communications and High Technologies, National Nuclear Research Center, Azerbaijan;
Conselho Nacional de Desenvolvimento Cient\'{\i}fico e Tecnol\'{o}gico (CNPq), Universidade Federal do Rio Grande do Sul (UFRGS), Financiadora de Estudos e Projetos (Finep) and Funda\c{c}\~{a}o de Amparo \`{a} Pesquisa do Estado de S\~{a}o Paulo (FAPESP), Brazil;
Ministry of Science \& Technology of China (MSTC), National Natural Science Foundation of China (NSFC) and Ministry of Education of China (MOEC) , China;
Ministry of Science and Education, Croatia;
Ministry of Education, Youth and Sports of the Czech Republic, Czech Republic;
The Danish Council for Independent Research | Natural Sciences, the Carlsberg Foundation and Danish National Research Foundation (DNRF), Denmark;
Helsinki Institute of Physics (HIP), Finland;
Commissariat \`{a} l'Energie Atomique (CEA) and Institut National de Physique Nucl\'{e}aire et de Physique des Particules (IN2P3) and Centre National de la Recherche Scientifique (CNRS), France;
Bundesministerium f\"{u}r Bildung, Wissenschaft, Forschung und Technologie (BMBF) and GSI Helmholtzzentrum f\"{u}r Schwerionenforschung GmbH, Germany;
General Secretariat for Research and Technology, Ministry of Education, Research and Religions, Greece;
National Research, Development and Innovation Office, Hungary;
Department of Atomic Energy Government of India (DAE), Department of Science and Technology, Government of India (DST), University Grants Commission, Government of India (UGC) and Council of Scientific and Industrial Research (CSIR), India;
Indonesian Institute of Science, Indonesia;
Centro Fermi - Museo Storico della Fisica e Centro Studi e Ricerche Enrico Fermi and Istituto Nazionale di Fisica Nucleare (INFN), Italy;
Institute for Innovative Science and Technology , Nagasaki Institute of Applied Science (IIST), Japan Society for the Promotion of Science (JSPS) KAKENHI and Japanese Ministry of Education, Culture, Sports, Science and Technology (MEXT), Japan;
Consejo Nacional de Ciencia (CONACYT) y Tecnolog\'{i}a, through Fondo de Cooperaci\'{o}n Internacional en Ciencia y Tecnolog\'{i}a (FONCICYT) and Direcci\'{o}n General de Asuntos del Personal Academico (DGAPA), Mexico;
Nederlandse Organisatie voor Wetenschappelijk Onderzoek (NWO), Netherlands;
The Research Council of Norway, Norway;
Commission on Science and Technology for Sustainable Development in the South (COMSATS), Pakistan;
Pontificia Universidad Cat\'{o}lica del Per\'{u}, Peru;
Ministry of Science and Higher Education and National Science Centre, Poland;
Korea Institute of Science and Technology Information and National Research Foundation of Korea (NRF), Republic of Korea;
Ministry of Education and Scientific Research, Institute of Atomic Physics and Romanian National Agency for Science, Technology and Innovation, Romania;
Joint Institute for Nuclear Research (JINR), Ministry of Education and Science of the Russian Federation and National Research Centre Kurchatov Institute, Russia;
Ministry of Education, Science, Research and Sport of the Slovak Republic, Slovakia;
National Research Foundation of South Africa, South Africa;
Centro de Aplicaciones Tecnol\'{o}gicas y Desarrollo Nuclear (CEADEN), Cubaenerg\'{\i}a, Cuba and Centro de Investigaciones Energ\'{e}ticas, Medioambientales y Tecnol\'{o}gicas (CIEMAT), Spain;
Swedish Research Council (VR) and Knut \& Alice Wallenberg Foundation (KAW), Sweden;
European Organization for Nuclear Research, Switzerland;
National Science and Technology Development Agency (NSDTA), Suranaree University of Technology (SUT) and Office of the Higher Education Commission under NRU project of Thailand, Thailand;
Turkish Atomic Energy Agency (TAEK), Turkey;
National Academy of  Sciences of Ukraine, Ukraine;
Science and Technology Facilities Council (STFC), United Kingdom;
National Science Foundation of the United States of America (NSF) and United States Department of Energy, Office of Nuclear Physics (DOE NP), United States of America.

%% file: Alice_Authorlist_2018-Apr-24.tex
% Collaboration: CERN-LHC-ALICE
% Generation Date is 2018-Apr-24

% How to use:
%%%%%%%%% appendix with author list
%\appendix
%\section{The ALICE Collaboration}
%\label{app:collab}
%\input{Alice_Authorslist_XXXX-Axx-XX.tex}
\begingroup
\small
\begin{flushleft}
S.~Acharya\Irefn{org138}\And 
F.T.-.~Acosta\Irefn{org22}\And 
D.~Adamov\'{a}\Irefn{org94}\And 
J.~Adolfsson\Irefn{org81}\And 
M.M.~Aggarwal\Irefn{org98}\And 
G.~Aglieri Rinella\Irefn{org36}\And 
M.~Agnello\Irefn{org33}\And 
N.~Agrawal\Irefn{org49}\And 
Z.~Ahammed\Irefn{org138}\And 
S.U.~Ahn\Irefn{org77}\And 
S.~Aiola\Irefn{org143}
% ...
\And
\renewcommand\labelenumi{\textsuperscript{\theenumi}~}

\section*{Affiliation notes}
\renewcommand\theenumi{\roman{enumi}}
\begin{Authlist}
\item \Adef{org*}Deceased
\item \Adef{orgI}Dipartimento DET del Politecnico di Torino, Turin, Italy
\item \Adef{orgII}M.V. Lomonosov Moscow State University, D.V. Skobeltsyn Institute of Nuclear, Physics, Moscow, Russia
\item \Adef{orgIII}Department of Applied Physics, Aligarh Muslim University, Aligarh, India
\item \Adef{orgIV}Institute of Theoretical Physics, University of Wroclaw, Poland
\end{Authlist}

\section*{Collaboration Institutes}
\renewcommand\theenumi{\arabic{enumi}~}
\begin{Authlist}
\item \Idef{org1}A.I. Alikhanyan National Science Laboratory (Yerevan Physics Institute) Foundation, Yerevan, Armenia
\item \Idef{org2}Benem\'{e}rita Universidad Aut\'{o}noma de Puebla, Puebla, Mexico
\item \Idef{org3}Bogolyubov Institute for Theoretical Physics, National Academy of Sciences of Ukraine, Kiev, Ukraine
\item \Idef{org4}Bose Institute, Department of Physics  and Centre for Astroparticle Physics and Space Science (CAPSS), Kolkata, India
\item \Idef{org5}Budker Institute for Nuclear Physics, Novosibirsk, Russia
\item \Idef{org6}California Polytechnic State University, San Luis Obispo, California, United States
\item \Idef{org7}Central China Normal University, Wuhan, China
\item \Idef{org8}Centre de Calcul de l'IN2P3, Villeurbanne, Lyon, France
\item \Idef{org9}Centro de Aplicaciones Tecnol\'{o}gicas y Desarrollo Nuclear (CEADEN), Havana, Cuba
\item \Idef{org10}Centro de Investigaci\'{o}n y de Estudios Avanzados (CINVESTAV), Mexico City and M\'{e}rida, Mexico
\item \Idef{org11}Centro Fermi - Museo Storico della Fisica e Centro Studi e Ricerche ``Enrico Fermi', Rome, Italy
\item \Idef{org12}Chicago State University, Chicago, Illinois, United States
\item \Idef{org13}China Institute of Atomic Energy, Beijing, China
\item \Idef{org14}Chonbuk National University, Jeonju, Republic of Korea
\item \Idef{org15}Comenius University Bratislava, Faculty of Mathematics, Physics and Informatics, Bratislava, Slovakia
\item \Idef{org16}COMSATS Institute of Information Technology (CIIT), Islamabad, Pakistan
\item \Idef{org17}Creighton University, Omaha, Nebraska, United States
\item \Idef{org18}Department of Physics, Aligarh Muslim University, Aligarh, India
\item \Idef{org19}Department of Physics, Ohio State University, Columbus, Ohio, United States
\item \Idef{org20}Department of Physics, Pusan National University, Pusan, Republic of Korea
\item \Idef{org21}Department of Physics, Sejong University, Seoul, Republic of Korea
\item \Idef{org22}Department of Physics, University of California, Berkeley, California, United States
\item \Idef{org23}Department of Physics, University of Oslo, Oslo, Norway
\item \Idef{org24}Department of Physics and Technology, University of Bergen, Bergen, Norway
\item \Idef{org25}Dipartimento di Fisica dell'Universit\`{a} 'La Sapienza' and Sezione INFN, Rome, Italy
\item \Idef{org26}Dipartimento di Fisica dell'Universit\`{a} and Sezione INFN, Cagliari, Italy
\item \Idef{org27}Dipartimento di Fisica dell'Universit\`{a} and Sezione INFN, Trieste, Italy
\item \Idef{org28}Dipartimento di Fisica dell'Universit\`{a} and Sezione INFN, Turin, Italy
\item \Idef{org29}Dipartimento di Fisica e Astronomia dell'Universit\`{a} and Sezione INFN, Bologna, Italy
\item \Idef{org30}Dipartimento di Fisica e Astronomia dell'Universit\`{a} and Sezione INFN, Catania, Italy
\item \Idef{org31}Dipartimento di Fisica e Astronomia dell'Universit\`{a} and Sezione INFN, Padova, Italy
\item \Idef{org32}Dipartimento di Fisica `E.R.~Caianiello' dell'Universit\`{a} and Gruppo Collegato INFN, Salerno, Italy
\item \Idef{org33}Dipartimento DISAT del Politecnico and Sezione INFN, Turin, Italy
\item \Idef{org34}Dipartimento di Scienze e Innovazione Tecnologica dell'Universit\`{a} del Piemonte Orientale and INFN Sezione di Torino, Alessandria, Italy
\item \Idef{org35}Dipartimento Interateneo di Fisica `M.~Merlin' and Sezione INFN, Bari, Italy
\item \Idef{org36}European Organization for Nuclear Research (CERN), Geneva, Switzerland
\item \Idef{org37}Faculty of Electrical Engineering, Mechanical Engineering and Naval Architecture, University of Split, Split, Croatia
\item \Idef{org38}Faculty of Engineering and Science, Western Norway University of Applied Sciences, Bergen, Norway
\item \Idef{org39}Faculty of Nuclear Sciences and Physical Engineering, Czech Technical University in Prague, Prague, Czech Republic
\item \Idef{org40}Faculty of Science, P.J.~\v{S}af\'{a}rik University, Ko\v{s}ice, Slovakia
\item \Idef{org41}Frankfurt Institute for Advanced Studies, Johann Wolfgang Goethe-Universit\"{a}t Frankfurt, Frankfurt, Germany
\item \Idef{org42}Gangneung-Wonju National University, Gangneung, Republic of Korea
\item \Idef{org43}Gauhati University, Department of Physics, Guwahati, India
\item \Idef{org44}Helmholtz-Institut f\"{u}r Strahlen- und Kernphysik, Rheinische Friedrich-Wilhelms-Universit\"{a}t Bonn, Bonn, Germany
\item \Idef{org45}Helsinki Institute of Physics (HIP), Helsinki, Finland
\item \Idef{org46}Hiroshima University, Hiroshima, Japan
\item \Idef{org47}Hochschule Worms, Zentrum  f\"{u}r Technologietransfer und Telekommunikation (ZTT), Worms, Germany
\item \Idef{org48}Horia Hulubei National Institute of Physics and Nuclear Engineering, Bucharest, Romania
\item \Idef{org49}Indian Institute of Technology Bombay (IIT), Mumbai, India
\item \Idef{org50}Indian Institute of Technology Indore, Indore, India
\item \Idef{org51}Indonesian Institute of Sciences, Jakarta, Indonesia
\item \Idef{org52}INFN, Laboratori Nazionali di Frascati, Frascati, Italy
\item \Idef{org53}INFN, Sezione di Bari, Bari, Italy
\item \Idef{org54}INFN, Sezione di Bologna, Bologna, Italy
\item \Idef{org55}INFN, Sezione di Cagliari, Cagliari, Italy
\item \Idef{org56}INFN, Sezione di Catania, Catania, Italy
\item \Idef{org57}INFN, Sezione di Padova, Padova, Italy
\item \Idef{org58}INFN, Sezione di Roma, Rome, Italy
\item \Idef{org59}INFN, Sezione di Torino, Turin, Italy
\item \Idef{org60}INFN, Sezione di Trieste, Trieste, Italy
\item \Idef{org61}Inha University, Incheon, Republic of Korea
\item \Idef{org62}Institut de Physique Nucl\'{e}aire d'Orsay (IPNO), Institut National de Physique Nucl\'{e}aire et de Physique des Particules (IN2P3/CNRS), Universit\'{e} de Paris-Sud, Universit\'{e} Paris-Saclay, Orsay, France
\item \Idef{org63}Institute for Nuclear Research, Academy of Sciences, Moscow, Russia
\item \Idef{org64}Institute for Subatomic Physics, Utrecht University/Nikhef, Utrecht, Netherlands
\item \Idef{org65}Institute for Theoretical and Experimental Physics, Moscow, Russia
\item \Idef{org66}Institute of Experimental Physics, Slovak Academy of Sciences, Ko\v{s}ice, Slovakia
\item \Idef{org67}Institute of Physics, Bhubaneswar, India
\item \Idef{org68}Institute of Physics of the Czech Academy of Sciences, Prague, Czech Republic
\item \Idef{org69}Institute of Space Science (ISS), Bucharest, Romania
\item \Idef{org70}Institut f\"{u}r Kernphysik, Johann Wolfgang Goethe-Universit\"{a}t Frankfurt, Frankfurt, Germany
\item \Idef{org71}Instituto de Ciencias Nucleares, Universidad Nacional Aut\'{o}noma de M\'{e}xico, Mexico City, Mexico
\item \Idef{org72}Instituto de F\'{i}sica, Universidade Federal do Rio Grande do Sul (UFRGS), Porto Alegre, Brazil
\item \Idef{org73}Instituto de F\'{\i}sica, Universidad Nacional Aut\'{o}noma de M\'{e}xico, Mexico City, Mexico
\item \Idef{org74}iThemba LABS, National Research Foundation, Somerset West, South Africa
\item \Idef{org75}Johann-Wolfgang-Goethe Universit\"{a}t Frankfurt Institut f\"{u}r Informatik, Fachbereich Informatik und Mathematik, Frankfurt, Germany
\item \Idef{org76}Joint Institute for Nuclear Research (JINR), Dubna, Russia
\item \Idef{org77}Korea Institute of Science and Technology Information, Daejeon, Republic of Korea
\item \Idef{org78}KTO Karatay University, Konya, Turkey
\item \Idef{org79}Laboratoire de Physique Subatomique et de Cosmologie, Universit\'{e} Grenoble-Alpes, CNRS-IN2P3, Grenoble, France
\item \Idef{org80}Lawrence Berkeley National Laboratory, Berkeley, California, United States
\item \Idef{org81}Lund University Department of Physics, Division of Particle Physics, Lund, Sweden
\item \Idef{org82}Nagasaki Institute of Applied Science, Nagasaki, Japan
\item \Idef{org83}Nara Women{'}s University (NWU), Nara, Japan
\item \Idef{org84}National and Kapodistrian University of Athens, School of Science, Department of Physics , Athens, Greece
\item \Idef{org85}National Centre for Nuclear Research, Warsaw, Poland
\item \Idef{org86}National Institute of Science Education and Research, HBNI, Jatni, India
\item \Idef{org87}National Nuclear Research Center, Baku, Azerbaijan
\item \Idef{org88}National Research Centre Kurchatov Institute, Moscow, Russia
\item \Idef{org89}Niels Bohr Institute, University of Copenhagen, Copenhagen, Denmark
\item \Idef{org90}Nikhef, National institute for subatomic physics, Amsterdam, Netherlands
\item \Idef{org91}NRC Kurchatov Institute IHEP, Protvino, Russia
\item \Idef{org92}NRNU Moscow Engineering Physics Institute, Moscow, Russia
\item \Idef{org93}Nuclear Physics Group, STFC Daresbury Laboratory, Daresbury, United Kingdom
\item \Idef{org94}Nuclear Physics Institute of the Czech Academy of Sciences, \v{R}e\v{z} u Prahy, Czech Republic
\item \Idef{org95}Oak Ridge National Laboratory, Oak Ridge, Tennessee, United States
\item \Idef{org96}Petersburg Nuclear Physics Institute, Gatchina, Russia
\item \Idef{org97}Physics department, Faculty of science, University of Zagreb, Zagreb, Croatia
\item \Idef{org98}Physics Department, Panjab University, Chandigarh, India
\item \Idef{org99}Physics Department, University of Jammu, Jammu, India
\item \Idef{org100}Physics Department, University of Rajasthan, Jaipur, India
\item \Idef{org101}Physikalisches Institut, Eberhard-Karls-Universit\"{a}t T\"{u}bingen, T\"{u}bingen, Germany
\item \Idef{org102}Physikalisches Institut, Ruprecht-Karls-Universit\"{a}t Heidelberg, Heidelberg, Germany
\item \Idef{org103}Physik Department, Technische Universit\"{a}t M\"{u}nchen, Munich, Germany
\item \Idef{org104}Research Division and ExtreMe Matter Institute EMMI, GSI Helmholtzzentrum f\"ur Schwerionenforschung GmbH, Darmstadt, Germany
\item \Idef{org105}Rudjer Bo\v{s}kovi\'{c} Institute, Zagreb, Croatia
\item \Idef{org106}Russian Federal Nuclear Center (VNIIEF), Sarov, Russia
\item \Idef{org107}Saha Institute of Nuclear Physics, Kolkata, India
\item \Idef{org108}School of Physics and Astronomy, University of Birmingham, Birmingham, United Kingdom
\item \Idef{org109}Secci\'{o}n F\'{\i}sica, Departamento de Ciencias, Pontificia Universidad Cat\'{o}lica del Per\'{u}, Lima, Peru
\item \Idef{org110}Shanghai Institute of Applied Physics, Shanghai, China
\item \Idef{org111}Stefan Meyer Institut f\"{u}r Subatomare Physik (SMI), Vienna, Austria
\item \Idef{org112}SUBATECH, IMT Atlantique, Universit\'{e} de Nantes, CNRS-IN2P3, Nantes, France
\item \Idef{org113}Suranaree University of Technology, Nakhon Ratchasima, Thailand
\item \Idef{org114}Technical University of Ko\v{s}ice, Ko\v{s}ice, Slovakia
\item \Idef{org115}Technische Universit\"{a}t M\"{u}nchen, Excellence Cluster 'Universe', Munich, Germany
\item \Idef{org116}The Henryk Niewodniczanski Institute of Nuclear Physics, Polish Academy of Sciences, Cracow, Poland
\item \Idef{org117}The University of Texas at Austin, Austin, Texas, United States
\item \Idef{org118}Universidad Aut\'{o}noma de Sinaloa, Culiac\'{a}n, Mexico
\item \Idef{org119}Universidade de S\~{a}o Paulo (USP), S\~{a}o Paulo, Brazil
\item \Idef{org120}Universidade Estadual de Campinas (UNICAMP), Campinas, Brazil
\item \Idef{org121}Universidade Federal do ABC, Santo Andre, Brazil
\item \Idef{org122}University College of Southeast Norway, Tonsberg, Norway
\item \Idef{org123}University of Cape Town, Cape Town, South Africa
\item \Idef{org124}University of Houston, Houston, Texas, United States
\item \Idef{org125}University of Jyv\"{a}skyl\"{a}, Jyv\"{a}skyl\"{a}, Finland
\item \Idef{org126}University of Liverpool, Department of Physics Oliver Lodge Laboratory , Liverpool, United Kingdom
\item \Idef{org127}University of Tennessee, Knoxville, Tennessee, United States
\item \Idef{org128}University of the Witwatersrand, Johannesburg, South Africa
\item \Idef{org129}University of Tokyo, Tokyo, Japan
\item \Idef{org130}University of Tsukuba, Tsukuba, Japan
\item \Idef{org131}Universit\'{e} Clermont Auvergne, CNRS/IN2P3, LPC, Clermont-Ferrand, France
\item \Idef{org132}Universit\'{e} de Lyon, Universit\'{e} Lyon 1, CNRS/IN2P3, IPN-Lyon, Villeurbanne, Lyon, France
\item \Idef{org133}Universit\'{e} de Strasbourg, CNRS, IPHC UMR 7178, F-67000 Strasbourg, France, Strasbourg, France
\item \Idef{org134} Universit\'{e} Paris-Saclay Centre d¿\'Etudes de Saclay (CEA), IRFU, Department de Physique Nucl\'{e}aire (DPhN), Saclay, France
\item \Idef{org135}Universit\`{a} degli Studi di Pavia, Pavia, Italy
\item \Idef{org136}Universit\`{a} di Brescia, Brescia, Italy
\item \Idef{org137}V.~Fock Institute for Physics, St. Petersburg State University, St. Petersburg, Russia
\item \Idef{org138}Variable Energy Cyclotron Centre, Kolkata, India
\item \Idef{org139}Warsaw University of Technology, Warsaw, Poland
\item \Idef{org140}Wayne State University, Detroit, Michigan, United States
\item \Idef{org141}Westf\"{a}lische Wilhelms-Universit\"{a}t M\"{u}nster, Institut f\"{u}r Kernphysik, M\"{u}nster, Germany
\item \Idef{org142}Wigner Research Centre for Physics, Hungarian Academy of Sciences, Budapest, Hungary
\item \Idef{org143}Yale University, New Haven, Connecticut, United States
\item \Idef{org144}Yonsei University, Seoul, Republic of Korea
\end{Authlist}
\endgroup